\documentclass[aps,preprint,prd,nofootinbib]{revtex4}

\usepackage{amsmath}
\usepackage{graphicx}
\usepackage[colorlinks,linkcolor=magenta,anchorcolor=cyan,citecolor=blue]{hyperref}
\usepackage{soul}

\def\be{\begin{equation}}
\def\ee{\end{equation}}
\def\ba{\begin{eqnarray}}
\def\ea{\end{eqnarray}}

\def\lf{\left}
\def\rt{\right}

\begin{document}

\title{Primordial extreme mass-ratio inspirals }

\author{Hai-Long Huang$^{1,2}$\footnote{huanghailong18@mails.ucas.ac.cn}}
\author{Tian-Yi Song$^{2}$\footnote{songtianyi20@mails.ucas.ac.cn}}
\author{Yun-Song Piao$^{1,2,3,4}$\footnote{yspiao@ucas.ac.cn}}

\affiliation{$^1$ School of Fundamental Physics and Mathematical
    Sciences, Hangzhou Institute for Advanced Study, UCAS, Hangzhou
    310024, China}

\affiliation{$^2$ School of Physical Sciences, University of
Chinese Academy of Sciences, Beijing 100049, China}


\affiliation{$^3$ International Center for Theoretical Physics
    Asia-Pacific, Beijing/Hangzhou, China}

\affiliation{$^4$ Institute of Theoretical Physics, Chinese
    Academy of Sciences, P.O. Box 2735, Beijing 100190, China}



\begin{abstract}

The coalescence of stellar-mass primordial black holes (PBHs)
might explain some of the gravitation waves (GWs) events detected
by LIGO-Virgo-KAGRA. On the other hand, observational hints for
supermassive PBHs (SMPBHs) have been accumulated. Thus it can be
expected that stellar-mass PBHs might be gravitationally bounded
to SMPBHs ($\sim10^{6}-10^9M_\odot$) in the early Universe, and
both constituted primordial extreme mass-ratio inspirals (EMRIs).
In this work, we initiate the study of the merger rate for
primordial EMRIs. The corresponding intrinsic EMRI rate at low
redshift may be comparable to that of astrophysical model,
$10-10^4$yr$^{-1}$, which the space-based detector LISA has the
capability to detect, but significantly raises with redshift.
Though equal-mass binaries also inevitably form, we find that
under certain conditions the primordial EMRIs can be the most
prevalent GW sources, and thus potentially a new probe to PBH.

\end{abstract}

\maketitle

\section{Introduction}
\label{sec:introduction}

The possibility that the coalescence of PBHs
\cite{Zeldovich:1967lct,Hawking:1971ei,Carr:1974nx} sourced the GW
events detected by LIGO-Virgo-KAGRA \cite{LIGOScientific:2016aoc,
LIGOScientific:2016sjg,LIGOScientific:2016dsl,LIGOScientific:2017bnn}
have been intensively studied, see e.g.,
Refs.~\cite{Sasaki:2018dmp,Carr:2020gox,Carr:2023tpt,Domenech:2024cjn}
for relevant reviews. On the other hand, observational hints for
SMPBHs have been gathered and will continue to accumulate. In
particular, a nano-Hertz SGWB discovered by PTA
\cite{NANOGrav:2023gor,Xu:2023wog,Reardon:2023gzh,EPTA:2023fyk}
might be interpreted with a population of SMPBHs ($M\sim
10^9M_\odot$), e.g.
\cite{Huang:2023chx,Huang:2023mwy,Depta:2023qst,Gouttenoire:2023nzr,Hooper:2023nnl}.
And the observations with JWST have discovered lots of
supermassive galaxies and BHs with $M\sim 10^6-10^{10}M_\odot$ at
high redshifts ($z\gtrsim 6$, even $z\gtrsim 10$), which might also
suggest the existence of SMPBHs.

The upcoming space-based detectors, e.g., LISA \cite{LISA:2017pwj}
and Taiji \cite{Hu:2017mde}, aim for detecting the merger of
supermassive BH binaries with $M\sim 10^4-10^7M_\odot$. However,
EMRI \cite{Amaro-Seoane:2007osp,LISA:2022yao} is also the
significant source for LISA, which refers to the systems comprised
of stellar-mass BHs ($M_1$) or other comparable mass compact
objects orbiting around a massive or supermassive BH ($M_2$),
thus a mass ratio is $10^{-8}<q=M_1/M_2<10^{-5}$. The
maximally achievable GW frequency of EMRI is approximately
\cite{Bardeen:1972fi,Maggiore:2018sht}
\begin{equation} \label{eq:ISCO}
    f_{\text{ISCO}}\simeq4.4\text{kHz}\lf(\frac{M_\odot}{M_1+M_2}\rt)
    \simeq4.4\text{kHz}\lf(\frac{M_\odot}{M_2}\rt),
\end{equation}
which occurs at the innermost stable circular orbit (ISCO). Thus we have
$f_{\text{ISCO}}\sim\mathcal{O}(1)\text{mHz}$ for $M_2\sim
10^{6}M_\odot$, which is just the most sensitive frequency of
LISA. It has been widely approved that the EMRIs are ideal signals
to construct detailed maps of the background spacetime of massive
BHs \cite{PhysRevD.52.5707,PhysRevD.56.1845,
PhysRevD.56.7732,Glampedakis:2005cf,Barack:2006pq}, see also e.g.
\cite{PhysRevD.69.082005,PhysRevD.79.084021,Arun:2008zn,Barack:2006pq,
Gair:2012nm,PhysRevD.75.064026,PhysRevD.77.104027,PhysRevD.83.044037,
PhysRevLett.107.171103,PhysRevD.89.104059,Barausse:2014pra,Berry:2019wgg}
for other relevant issues.

It is usually thought that EMRIs can born in a varity of
interesting astrophysical environment (both BHs are astrophysical
BHs, i.e. ABH+ABH), including two-body relaxation in nuclear star
clusters \cite{Sigurdsson:1996uz,Bar-Or:2014ika,
Amaro-Seoane:2012jcd}, tidal separation of binary stellar-mass BHs
\cite{ColemanMiller:2005rm}, capture of cores of giants
\cite{Davies:2005tc,DiStefano:2001ci}, and capture of BHs in
accretion discs around the massive BH \cite{Amaro-Seoane:2007osp}.
Recently, it has been also showed that some stellar-mass or
smaller PBH (see e.g.
Refs.~\cite{Inomata:2017okj,Kannike:2017bxn,Cheng:2018yyr,Lin:2020goi,
Ashoorioon:2020hln,Kawai:2021edk,Karam:2022nym,
Papanikolaou:2022did,Fu:2022ssq,Garcia-Bellido:2017mdw,Germani:2017bcs,
Byrnes:2018txb,Fu:2019ttf,Fu:2019vqc,Fu:2020lob,Zhai:2023azx,Chen:2024gqn,
Ragavendra:2020sop,Fu:2022ypp,Di:2017ndc,Motohashi:2017kbs,Yi:2020cut,
Ballesteros:2018wlw,Kamenshchik:2018sig,Qiu:2022klm,
Nakama:2018utx,2022ChPhC..46d5103Z,Gao:2020tsa,Kawana:2021tde,
Lin:2021vwc,Pi:2021dft,
Domenech:2021wkk,DeLuca:2022bjs,Kawaguchi:2023mgk,Choudhury:2023hfm,
Choudhury:2023fwk,Domenech:2023dxx,Pi:2022zxs,Domenech:2024rks,Flores:2023zpf,Flores:2024lng,
Cai:2023uhc,Choudhury:2023jlt,Choudhury:2023hvf,Bhattacharya:2023ysp,
Choudhury:2013woa,Wang:2023ost,Zhao:2022kvz,Li:2023xtl} for recent
studies) may accumulate at the center of a galaxy and constitute
EMRIs with central astrophysical massive or supermassive BH (i.e.
PBH+ABH)
\cite{Guo:2017njn,Kuhnel:2018mlr,Wang:2019kzb,Barsanti:2021ydd}.
However, since both stellar-mass PBHs and SMPBHs might be
ubiquitous at higher redshifts $z\gtrsim {\cal O}(10)$ where there
was little astrophysical competition, EMRIs are likely to come
into being earlier. It can be anticipated that unlike
astrophysical EMRIs, such primordial EMRIs (PBH+PBH) would not
only offer a more powerful tool to study the evolution of our
Universe but also potentially open a new avenue to identify PBHs.

In this paper, we carry out the first study of the merger rate for
primordial EMRIs, i.e. stellar-mass PBHs were gravitationally
bound to SMPBHs ($\sim10^6-10^9M_\odot$) in the early Universe,
and show its potential implications, in particular possibility as
a new probe to PBHs. Throughout this paper the values of
cosmological parameters are set in light of the Planck results
\cite{Planck:2018vyg}. The scale factor is normalized to unity at
the matter-radiation equality $z=z_\text{eq}\approx3400$ and we
denote by $t_0$ the present time.


\section{Modelling primordial EMRIs}
\label{sec:EMRIform}

The PBHs might have the normalized multi-peaks mass distribution
(multiple populations of PBHs with different masses $M_i$)
\begin{equation}
    \psi(m)=\sum_if_i\psi_{i}\left(m|\sigma_i,M_i\right) \quad \text{with}
    \quad \sum_if_i=1,
    \label{eq:twopeaks_bubble}
\end{equation}
with $\sigma_i$ and $M_i$ the width and the characteristic mass of
the $i$th peak, respectively. Here, we set
$\psi_i(m)=\frac{m}{\rho_{\text{PBH}}(M_i)}\frac{\text{d}n}{\text{d}m}$
with $\int\psi_i(m)\text{d}m=1$, so $\int\psi(m)\text{d}m=1$,
e.g.~\cite{Raidal:2018bbj,Hall:2020daa,Franciolini:2022tfm}.

The PBH model in Refs.~\cite{Huang:2023chx,Huang:2023klk} (inspired
by the seminal Refs.~\cite{Garriga:2015fdk,Sato:1981gv,Kodama:1982sf})
is an example for mass spectrum \autoref{eq:twopeaks_bubble}. In
corresponding model, the mass distribution of PBHs sourced by
supercritical bubbles that nucleated during inflation has a
peak-like spectrum \cite{Huang:2023klk},
\begin{equation}
\label{eq:peaks_bubble}
\psi_i(m|\sigma_i,M_i)=e^{-\sigma_i^2/8}\sqrt{\frac{M_i}{2\pi\sigma_i^2m^3}}
    \exp\lf(-\frac{\ln^2(m/M_i)}{2\sigma_i^2}\rt),
\end{equation}
with the characteristic mass $M_i<10^{18}M_\odot$ \cite{Huang:2023mwy}.
However, the slow-roll path of inflaton might be
accompanied with more than one neighboring vacua, as a result the
corresponding PBHs would present a normalized multi-peaks mass
distribution at different mass bands below $10^{18}M_\odot$,
\begin{equation}
    \psi(m)=\sum_if_i e^{-\sigma_i^2/8}\sqrt{\frac{M_i}{2\pi\sigma_i^2m^3}}
    \exp\lf(-\frac{\ln^2(m/M_i)}{2\sigma_i^2}\rt) \quad \text{with}
    \quad \sum_if_i=1,
    \label{eq:multipeaks_bubble}
\end{equation}
see also \cite{Huang:2023mwy} for possible clustering of PBHs.
Thus both stellar-mass PBHs and SMPBHs ($\sim10^6-10^9M_\odot$) might
coexist in the early Universe.

\begin{figure}[h!]
\centering
\includegraphics[width=1.\textwidth]{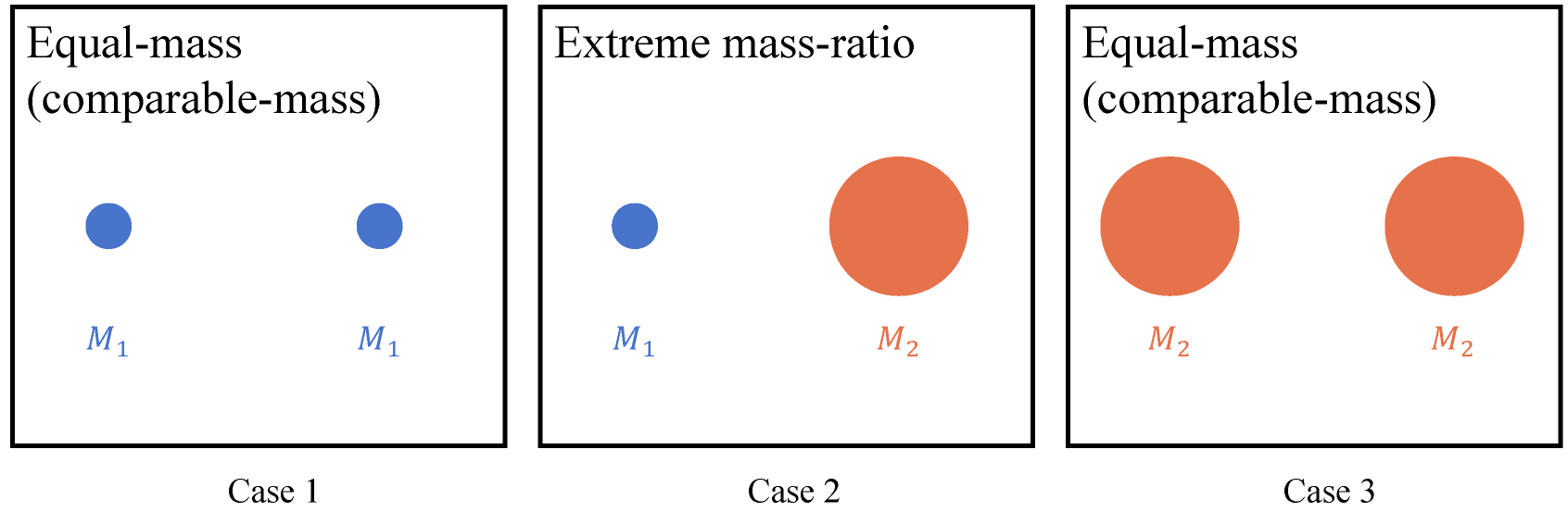}
\caption{\label{fig:binary} A schematic picture of possible PBH
binaries in a PBH model with a double-peaks mass function
\autoref{eq:twopeaksMF} in the radiation dominated epoch. }
\end{figure}

It is obvious in \autoref{eq:peaks_bubble} that $\psi_i$
approaches the monochromatic spectrum centered on $M_i$ as
$\sigma_i\to0$, $\psi_i(m|\sigma_i,M_i)\approx\delta(m-M_i)$. Here, for
our purpose, we focus on the scenario, in which only only two
monochromatic populations of PBHs with a large characteristic mass
ratio ($q=M_1/M_2\ll1$) exist. In such a scenario, we have
\begin{equation} \label{eq:twopeaksMF}
    \psi(m)=\sum_{i=1,2}f_i\delta(m-M_i) \quad \text{with}
    \quad f_1+f_2=1,
\end{equation}
where $f_i=\frac{\rho_{\text{PBH}}(M_i)}{\rho_{\text{PBH}}}$ is
the energy fraction of PBHs with mass $M_i$.
\autoref{eq:twopeaksMF} serves as a sufficiently good
approximation when the width of peak is far smaller than
$|M_1-M_2|$ (which is the case we are concerned about). And the
mean mass of PBHs is
\begin{equation}
    \lf\langle m\rt\rangle\equiv\frac{1}{n}\int m\text{d}n=
    \lf(\int\text{d}m\frac{\psi(m)}
    {m}\rt)^{-1}=\frac{M_1M_2}{f_1M_2+f_2M_1}
\end{equation}
with $n=\int {\text{d}}n(m)$, where the average number density of PBHs in
the mass interval $(m,m+\text{d}m)$ is
$n(m)\text{d}m=\frac{\rho_\text{PBH}\psi(m){\text{d}}m}{m}$. In
\autoref{fig:binary}, we present the possible binaries patterns,
one of which is EMRI. Thus EMRI can have a primordial origin.

\section{Merger rate of primordial EMRIs}

Initially, the different population of PBHs are randomly
distributed in the early Universe. Generally, when two PBHs with
masses of $m_i$ and $m_j$ happened to be separated with
\cite{Huang:2023klk}
\begin{equation} \label{eq:xmax}
    x<x_\text{max}=\lf(\frac{3}{8\pi}\cdot\frac{m_i+m_j}
    {\rho_\text{eq}}\rt)^{1/3},
\end{equation}
where $\rho_{\text{eq}}$ is the matter density at
$z_\text{eq}$, they will decouple from the Hubble flow. After
that, due to the disturbance of the environment, most of all being
a third neighboring BH exerting a tidal torque on them, they will
start inspiraling \cite{Kocsis:2017yty,Sasaki:2016jop,
Nakamura:1997sm,Ioka:1998nz}.

In our case, the mass of PBHs extends over many orders of
magnitude, thus the torques by all PBHs and linear density
perturbations must be taken into account,
e.g.~\cite{Ali-Haimoud:2017rtz,Chen:2018czv,Liu:2018ess,Huang:2023klk}.
According to \cite{Huang:2023klk}, we have
\begin{equation} \label{eq:a}
    a=\frac{0.1x^4}{x_\text{max}^3}=\frac{0.1\Bar{x}^4}
    {x_\text{max}^3}X^{4/3}
\end{equation}
is the semi-major axis $a$ of the binary orbit, with $X\equiv ({x/
\Bar{x}})^3$ the rescaled separation and $\Bar{x}=(4\pi
n_T/3)^{-1/3}$ the characteristic comoving separation between
nearest PBHs. Here, $n_T={\rho_{\text{PBH}}/\lf\langle m\rt\rangle}$
is the average number density of PBHs. And the probability distribution
of the dimensionless angular momentum $j$ for a given $X$ is
\begin{equation} \label{eq:jdPdj}
    P(j|X)=\frac{jj_X}{(j^2+j_X^2)^{3/2}} \quad \text{with} \quad
    j_X\approx\frac{\lf\langle m\rt\rangle}{m_i+m_j}\lf(1+\frac
    {\sigma_\text{eq}^2}{f^2}\rt)^{1/2}X,
\end{equation}
where $\sigma_{\text{eq}}=0.005$ is the variance of density
perturbations of the rest of dark matter at matter radiation
equality, and $f\approx0.85f_\text{PBH}$ is the abundance of PBHs in
non-relativistic matter.

The effect of cosmic expansion on the comoving separation of PBH
pair, so the merger rate, might be not negligible, when the number
density of PBHs is very low. It is actually this case for SMPBHs.
The current observations requires the fraction of PBHs
$f_\text{PBH}\lesssim 10^{-3}$ for $10^6M_\odot\lesssim M\lesssim
10^{12}M_\odot$ \cite{Carr:2023tpt,Carr:2020gox,Carr:2018rid},
thus the number density $\sim \rho_\text{PBH}f_2M_2^{-1}$ of SMPBHs is
actually considerably low.

The corresponding merger rate of PBHs is
\begin{equation}
    R(t)=\frac{1}{2}\frac{n_T}{(1+z_
    \text{eq})^3}\frac{\text{d}P}{\text{d}t}\equiv \int\int\mathcal{R}(m_i,m_j,t)\text{d}m_i\text{d}m_j,
\label{eq:merger}
\end{equation}
where $\mathcal{R}(m_i,m_j,t)$ is the differential merger rate,
${\text{d}P}/{\text{d}t}$ is the probability distribution of the
merger time. The merger rate of PBHs has
been studied in Refs.~\cite{Sasaki:2016jop,Ali-Haimoud:2017rtz,
Chen:2018czv,Liu:2018ess,Kocsis:2017yty,Raidal:2018bbj,Huang:2023klk},
in particular,
Refs.~\cite{Ali-Haimoud:2017rtz,Chen:2018czv,Liu:2018ess,Huang:2023klk}
considered the torques by all PBHs and linear density
perturbations. Taking further account for the effect of cosmic
expansion, we have \cite{Huang:2023klk}
\begin{align} \label{eq:lastdpdt}
    \frac{\text{d}P(m_i,m_j,X)}{\text{d}t}
    =\frac{1}{7t}\psi(m_i)\frac{\lf\langle m\rt\rangle}{m_i}\text{d}m_i\psi(m_j)\frac{\lf\langle m\rt\rangle}{m_j}
    \text{d}m_j\int\text{d}Xe^{-X}
    \Theta(X_\text{max}-X)\mathcal{P}(\gamma_X),
\end{align}
where $X_\text{max}=(x_\text{max}/\Bar{x})^3$, and
$\mathcal{P}(\gamma_X)\equiv\frac{\gamma_X^2}{\lf(1+\gamma_X^2\rt)^{3/2}}$
with
\begin{align}
    \gamma_X\approx10^{-3}f^{\frac{16}{21}}\lf(1+\frac{\sigma_{\text{eq}}^2}{f^2}\rt)^
    {-\frac{1}{2}}\lf(\frac{M_i}{M_\odot}\rt)^{\frac{1}{7}}\lf(\frac{M_j}
    {M_\odot}\rt)^{\frac{1}{7}}\lf(\frac{M_i+M_j}{M_\odot}\rt)^{\frac{12}{7}}
    \lf(\frac{\langle m\rangle}{M_\odot}\rt)^{-\frac{37}{21}}
    \lf(\frac{t}{t_0}\rt)^{\frac{1}{7}}X^{-\frac{37}{21}}.
\end{align}
Here, $\Theta(X_{\text{max}}-X)$ reflects the impact of cosmic
expansion on binding PBH binaries \cite{Huang:2023klk}.




In the case with only two discrete populations of PBHs, see
\autoref{eq:twopeaksMF}, we have
\begin{equation}\label{eq:total} R(t)= \sum_{ij} R_{ij}(t)\end{equation}
with $R_{ij}=R_{ji}$, where
\begin{align} \label{eq:mergerrate_allPBHs}
    R_{ij}&\approx\frac{2.04\times10^8}{\text{Gpc}^3\text{yr}}f_if_j
    f\lf(\frac{M_i}{M_\odot}\rt)^{-1}\lf(\frac{M_j}{M_\odot}\rt)^{-1}
    \lf(\frac{\lf\langle m\rt\rangle}{M_\odot}\rt)
    \lf(\frac{t}{t_0}\rt)^{-1}Y(M_i,M_j,t)
\end{align}
corresponds to the merger rate density $R_{ij}$, and
$Y(M_i,M_j,t)\equiv\int_0^{X_\text{max}}\text{d}Xe^{-X}\mathcal{P}(\gamma_X)$.
Here, $R_{11}$, $R_{12}+R_{21}$ and $R_{22}$ is nothing but the
merger rates of Case 1, Case 2 and Case 3, respectively, as
illustrated in \autoref{fig:binary}. The merger rate of primordial
EMRIs is $R_{12}+R_{21}$.

\section{Results and implications}
\label{sec:results}

It is well-known that EMRI is the significant source for LISA.
Here, we choose $M_2=10^6M_\odot$ with
$f_\text{PBH}f_2\lesssim10^{-3}$, compatible with current
constrains on the abundance of SMPBHs
\cite{Carr:2023tpt,Carr:2020gox,Carr:2018rid}. As a supplement, we
also consider $M_2=10^9M_\odot$. The GW events detected by
LIGO-Virgo-KAGRA \cite{LIGOScientific:2016aoc,LIGOScientific:2016sjg,LIGOScientific:2016dsl,
LIGOScientific:2017bnn}
imply the existence of BHs with masses $\sim 10M_\odot$ (the
merger rate of BH binaries inferred is $12-213$
$\text{Gpc}^{-3}\text{yr}^{-1}$ \cite{LIGOScientific:2017bnn}),
which might be (or in part) primordial. Therefore, it is natural
to choose $M_1=10M_\odot$.

\begin{figure}[h!]
\centering
\includegraphics[width=1.\textwidth]{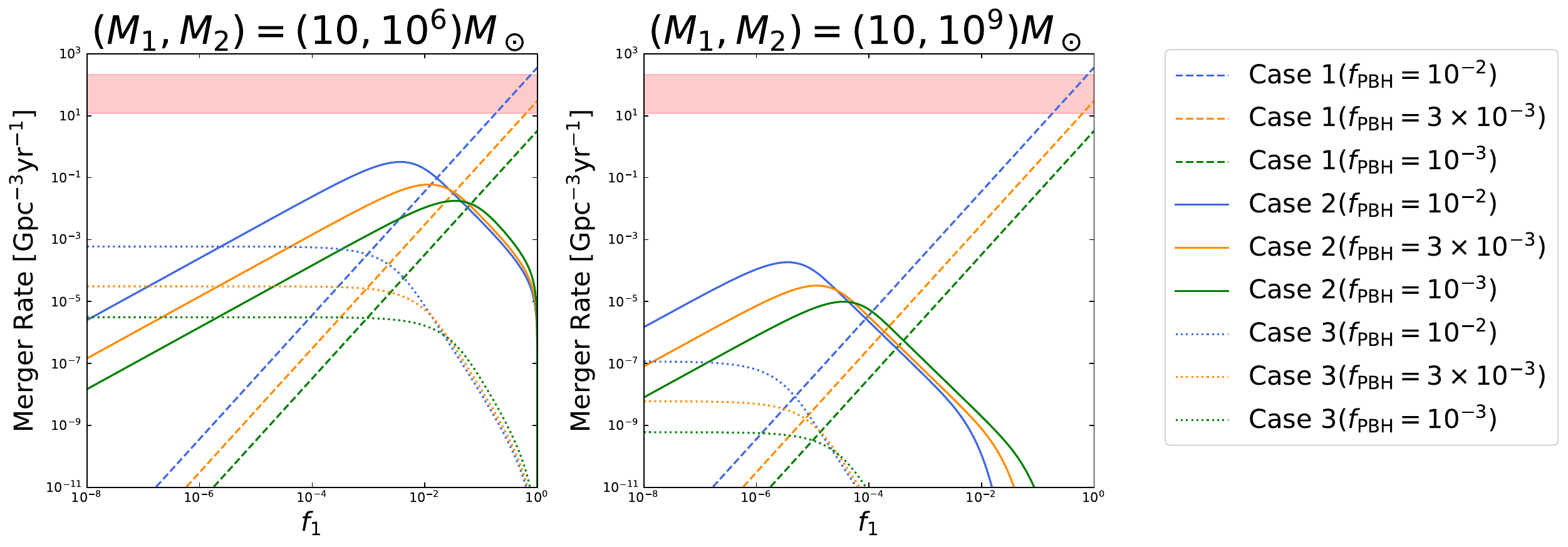}
\caption{\label{fig:mergerate_total} The merger rate for different
cases at redshift $z=0$ as a function of $f_1$ with different
$f_\text{PBH}$. The merger rate
$R_{11}=12-213~\text{Gpc}^{-3}\text{yr}^{-1}$ inferred by the LIGO
collaboration is shown as the shaded region colored red.
}
\end{figure}

According to \autoref{eq:mergerrate_allPBHs}, we can numerically
solve the merger rates. In \autoref{fig:mergerate_total}, we show
the merger rates for Case 1 ($R_{11}$), Case 2 ($R_{12}+R_{21}$),
and Case 3 ($R_{22}$) at $t=t_0$ with respect to $f_1$ for
different $f_\text{PBH}$. It is significantly found that the
primordial EMRI can be the most prevalent source when $f_1\sim
0.01$ ($M_2=10^{6}M_\odot$) or $10^{-5}$ ($M_2=10^{9}M_\odot$).

\begin{figure}[t]
{\includegraphics[width=8cm]{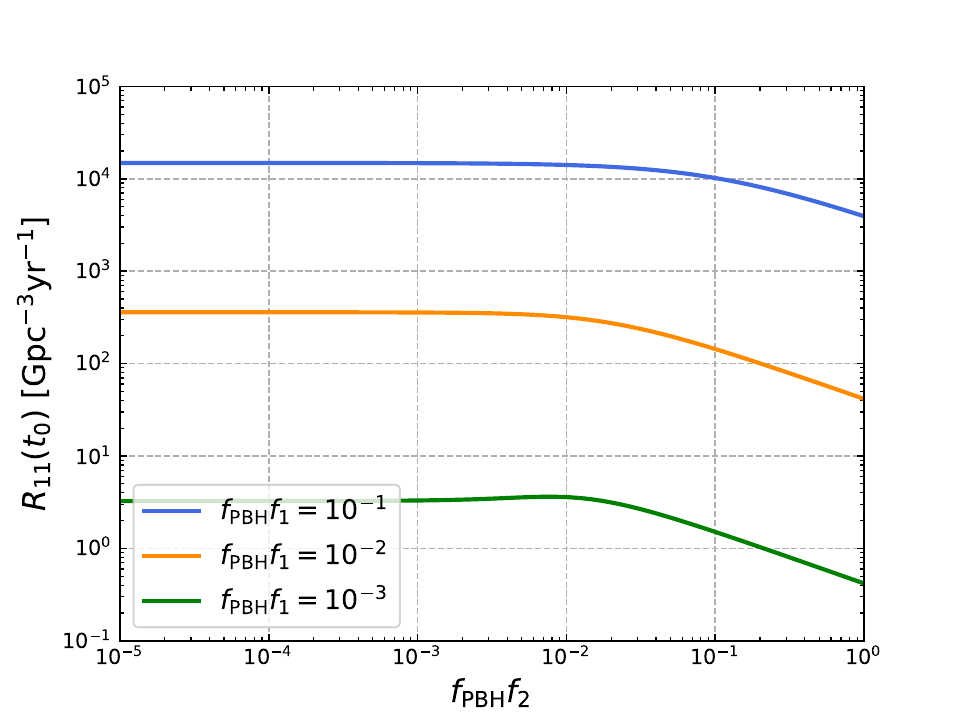}}
    { }
{\includegraphics[width=8cm]{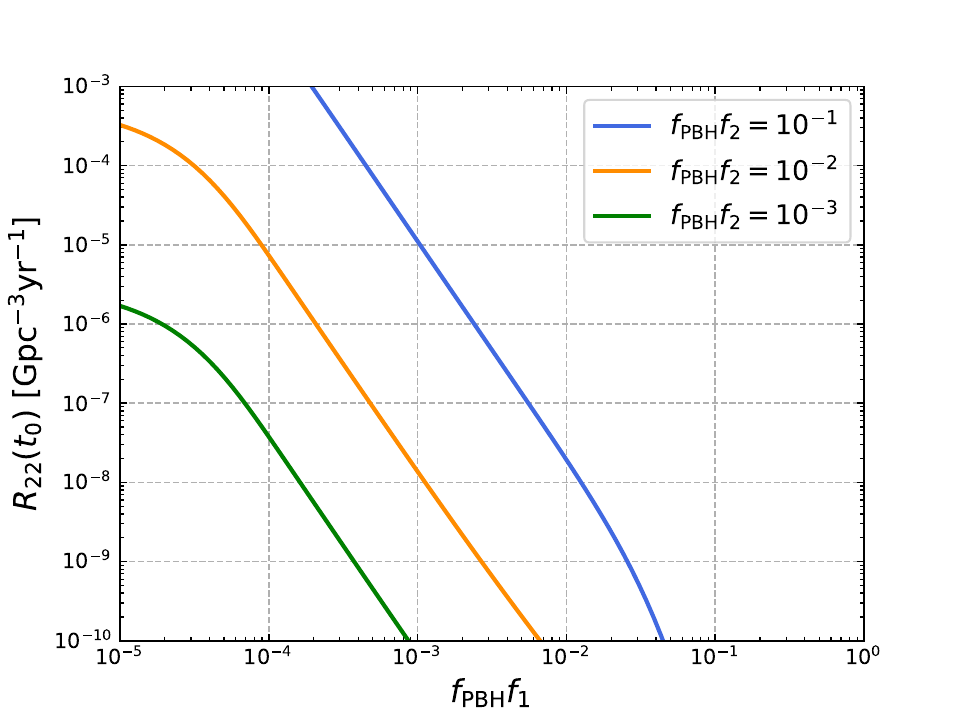}} \caption{The
effect of the presence of SMPBHs (stellar-mass BHs) on the merge
rate of equal-mass binaries $R_{11}$ ($R_{22}$). }
\label{fig:big_small}
\end{figure}

It is also observed that the presence of SMPBHs has a negligible
effect on the merge rate $R_{11}$ of stellar-mass BHs, except when
their energy density $f_{\text{PBH}}f_2$ exceeds that of
stellar-mass BHs, while since the number of stellar-mass BHs is
far larger than that of SMPBHs when stellar-mass BHs and SMPBHs
have the same energy density, the merger rate $R_{22}$ will be
strongly affected by $f_{\text{PBH}}f_1$. We illustrated it in
\autoref{fig:big_small}.


\begin{figure}[h!]
\centering
\includegraphics[width=1.\textwidth]{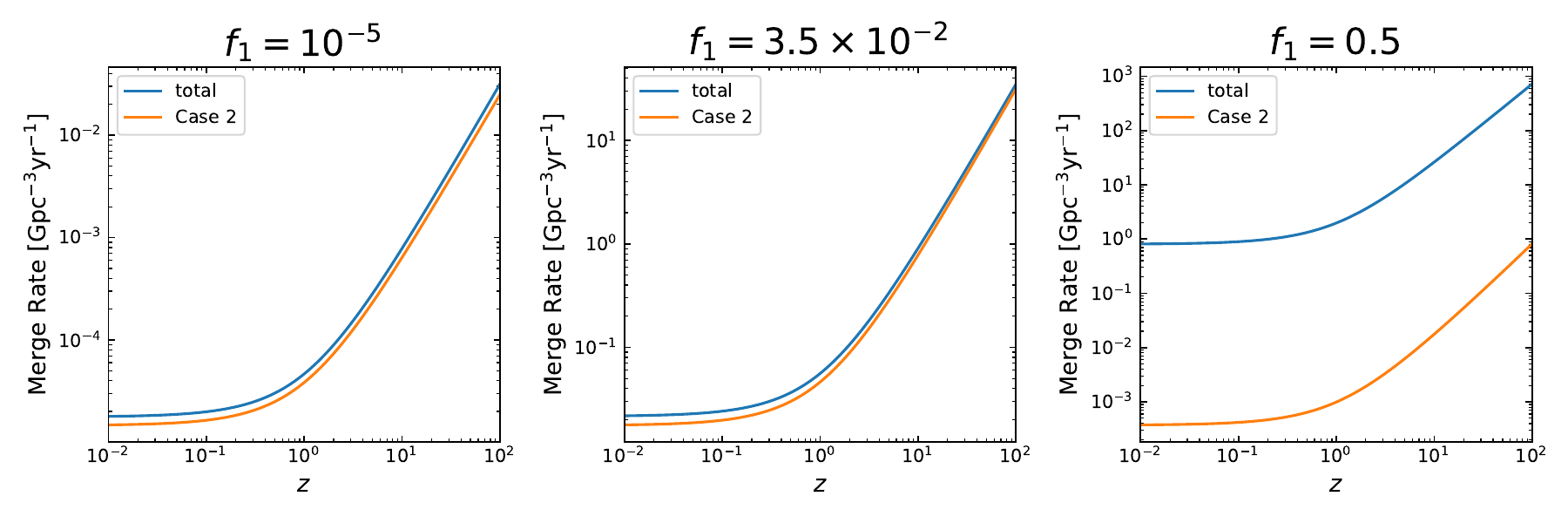}
\caption{\label{fig:mergerate_difff1} The merger rate
$R_{12}+R_{21}$ for EMRIs and the total merger rate
\autoref{eq:total} as a function of redshift with different $f_1$
($f_1=10^{-5}$ corresponds that the number density of the two
populations of PBHs is equal, $f_1=3.5\times10^{-2}$ corresponds
that the merger rate for EMRIs is maximized, see
\autoref{fig:mergerate_total}, and $f_1=0.5$ corresponds that the
energy density of the two populations of PBHs is equal), where
$f_\text{PBH}$ is fixed at $10^{-3}$. }
\end{figure}

In e.g. Ref.~\cite{Babak:2017tow}, astrophysically motivated EMRI
models have been presented, in which the uncertainties of the
number of EMRIs detectable by LISA are quantified,
see \autoref{tab:tab1} in Appendix \ref{sec:appendixB}. The
astrophysical EMRIs are thought to be cosmologically nearby, in
which case their merger rate is independent of redshift (or starts
to rapidly drop above $z\sim\mathcal{O}(1)$) within the
investigated range. However, unlike the astrophysical EMRIs, the
merger rate of primordial EMRIs monotonically raises with $z$, as
shown in \autoref{fig:mergerate_difff1}. In particular, for
$f_1=3.5\times10^{-2}$, the merger rate at $z\simeq 10$ can be
$>\mathcal{O}(1)\text{Gpc}^{-3}\text{yr}^{-1}$. Thus the evolution
of EMRI rate at high redshift $z\gtrsim\mathcal{O}(10)$ can be a
probe to PBHs.

\begin{figure}[h!]
\centering
\includegraphics[width=0.7\textwidth]{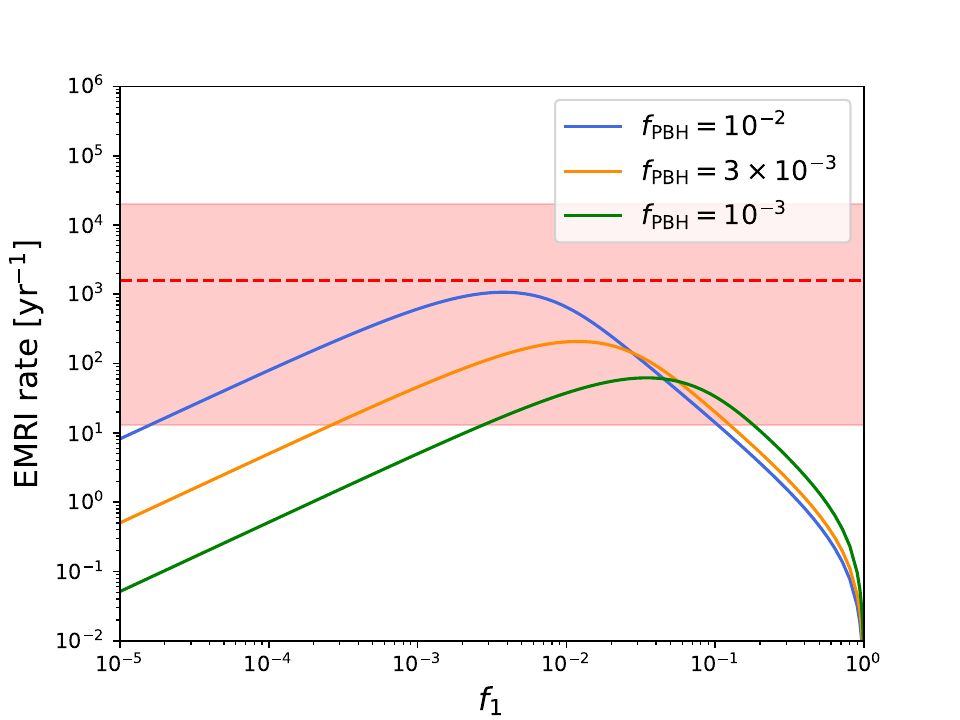}
\caption{\label{fig:abh_pbh} The intrinsic EMRI rate as a function
of our model parameters $f_1$ with different $f_{\text{PBH}}$
fixed, for $(M_1,M_2)=(10,10^6)M_\odot$. The shaded region
represent the potential range of astrophysical EMRI rates
calculated in Ref.~\cite{Babak:2017tow}. The red dashed line
corresponds to that of fiducial model M1 shown in
\autoref{tab:tab1}. }
\end{figure}

To compare the result with that of astrophysical EMRIs, it is
necessary to estimate the total intrinsic primordial EMRI rate (up
to $z=4.5$)
\begin{equation}
    r =\int_0^{z=4.5}\left(R_{12}+R_{21}\right)\frac{\text{d}V_c(z)}
    {\text{d}z}\frac{1}{1+z}\text{d}z.
\end{equation}
The intrinsic event rate of astrophysical EMRIs is
$10-10^4$yr$^{-1}$ \cite{Babak:2017tow}. As seen in
\autoref{fig:abh_pbh}, the resulting primordial EMRI rate is
comparable to that of astrophysical EMRI, suggesting that our
primordial channel could play an important role in LISA mission to
detect EMRIs.

\section{Discussion}
\label{sec:discussion}

In this paper, we present the primordial EMRIs and their merger
rate. In our scenario, stellar-mass PBHs were gravitationally
bound to SMPBHs ($\sim10^6-10^9M_\odot$). It is found that under
certain conditions the primordial EMRI can be the most prevalent
GW sources when $f_1\sim 0.01$ ($M_2=10^{6}M_\odot$) or $10^{-5}$
($M_2=10^{9}M_\odot$). And we also find that the corresponding
merger rate significantly raises with redshift, while the intrinsic
EMRI rate at low redshift is comparable to that of astrophysical model,
$10-10^4$yr$^{-1}$, which might be detected by LISA. Thus the
detection of such a EMRI merger at higher redshift will be
potentially a hint of PBHs.

There are much open issues left. It is relevant to compute the
possible detection rate of primordial EMRIs using LISA sensitivity
and compare it with the astrophysical EMRIs, see e.g.
\cite{Franciolini:2021xbq,LISACosmologyWorkingGroup:2023njw}. It
would be also very interesting for us to investigate the SGWB from
unresolved primordial EMRIs, see e.g.
Refs.~\cite{Barack:2004wc,Babak:2017tow,Bonetti:2020jku,
Pozzoli:2023kxy} for related researches on astrophysical EMRI
models. In our calculation, we neglected the spin of PBHs, and
also for the sake of simplicity we disregard the accretion and
clustering of PBHs during lengthy cosmic evolution (see e.g.
Refs.~\cite{DeLuca:2020qqa,DeLuca:2023bcr,DeLuca:2020jug}), it is
significant to assess the relevant effects. Here, the same
mechanism can also yield intermediate mass-ratio inspirals (IMRIs,
$10^{-5}<q<10^{-2}$) \cite{Amaro-Seoane:2007osp} and extremely
large mass-ratio inspirals (XMRIs, $q<10^{-8}$) e.g.
\cite{Rubbo:2006vh,Amaro-Seoane:2019umn}, as well as
binary-extreme mass-ratio inspiral (b-EMRI) e.g.
\cite{Addison:2019tib,Chen:2018axp}, however, it can be expected
that the primordial origin of relevant scenarios might bring
richer information of early Universe.

\section*{Acknowledgments}

We thanks Jun Zhang, Huai-Ke Guo, Yu-Tang Wang, Hao-Yang Liu and
Jibin He for helpful discussion. This work is supported by NSFC,
No.12075246, National Key Research and Development Program of
China, No. 2021YFC2203004, and the Fundamental Research Funds for
the Central Universities.

\appendix

\section{Direct plunge} \label{subsec:DirectPlunge}

It is known that the orbit of PBH binary formed in the early
Universe has a high eccentricity $e=\sqrt{1-j^2}$
\cite{Sasaki:2016jop, Nakamura:1997sm,Ioka:1998nz}. In the
modelling of primordial EMRIs, PBHs can also be swallowed whole if
they are kicked directly through the horizon (\textsf{direct
plunges}) in addition to inspiral gradually due to the emission of
GWs, as for astrophysical EMRIs \cite{Amaro-Seoane:2007osp}. This
happens when the pericentre distance is less than the
Schwarzschild radius of $M_2$, i.e.
\begin{equation} \label{eq:rprs}
    R_p=a(1-e)<R_s=2GM_2.
\end{equation}

Two BHs will eventually merge into one with the coalescence time
\cite{Peters:1964zz}
\begin{equation} \label{eq:peters}
    t(a,j)=\frac{3}{85}\frac{a^4}{G^3m_im_j(m_i+m_j)}j^7.
\end{equation}
According to \autoref{eq:a} and \autoref{eq:peters}, we have
\begin{align} \label{eq:x_jt}
    X(j,t)&={}\lf[\frac{85}{3}G^3m_im_j(m_i+m_j)\rt]^{3/16}\lf(\frac{0.1\Bar{x}^4}
    {x_\text{max}^3}\rt)^{-3/4}t^{3/16}j^{-21/16} \notag \\
    &\approx1.15\times10^{-4}f\frac{M_\odot}{\langle m\rangle}\lf[
    \frac{m_i^3m_j^3(m_i+m_j)^{15}}{M_\odot^{21}}\rt]^{1/16}\lf(\frac{t}{t_0}\rt)
    ^{3/16}j^{-21/16}.
\end{align}
Thus the probability distribution of $(X,j)$ is
\begin{equation}
    P(X,j)
    =e^{-X}\frac{jj_X}{\lf(j^2+j_X^2\rt)^{3/2}},
\end{equation}
where the probability distribution of the rescaled
nearest-neighbor separation is $P(X)=e^{-X}$ for a Poisson
distribution of PBHs.
The probability distribution of $j$ for the PBH binaries merging
after $t$ is
\begin{equation}
    P(j|t)=\alpha(t)\lf.P(X,j)\rt|_t=\alpha(t)\lf.\frac{e^{-X}jj_X}{\lf(j^2+j_X^2\rt)^{3/2}}\rt|_t,
\end{equation}
where $\alpha(t)$ is a normalization factor. The limitation on the
rescaled separation $X<X_\text{max}$ corresponds to
\begin{equation}
    j>j_\text{min}=(0.1)^{-4/7}\lf(\frac{8\pi\rho_\text{eq}}{3}\rt)^{4/21}
    \lf(\frac{85}{3}G^3m_im_j\rt)^{1/7}(m_i+m_j)^{-1/21}t^{1/7},
\end{equation}
which is independent on $\langle m\rangle$ or $\psi(m)$. On the
other hand, physical values are limited to $j\le1$. According to
\cite{Ali-Haimoud:2017rtz}, the contribution of unphysical values
$j>1$ is negligibly small as long as $j_X\ll1$.

In primordial EMRIs model, we replace $m_i$ and $m_j$ with $M_1$
and $M_2$ respectively. The characteristic value of semi-major
axis of primordial EMRIs is
\begin{equation}
    \Bar{a}=\frac{0.1\Bar{x}^4}{x_\text{max}^3}\approx1.07
    \times10^{-2}f^{-\frac{4}{3}}
    \lf(\frac{\langle m\rangle}{M_\odot}\rt)^{\frac{4}{3}}\lf
    (\frac{M_2}{M_\odot}\rt)^{-1}\text{pc}.
\end{equation}
Then we have $R_p<R_s$ only when $j<j_c$ in the case of
$a=\Bar{a}$, with
\begin{equation}
    j_c=\sqrt{1-\lf(1-\frac{2GM_2}{\Bar{a}}\rt)^2}.
\end{equation}

\begin{figure}[t]
{\includegraphics[width=8cm]{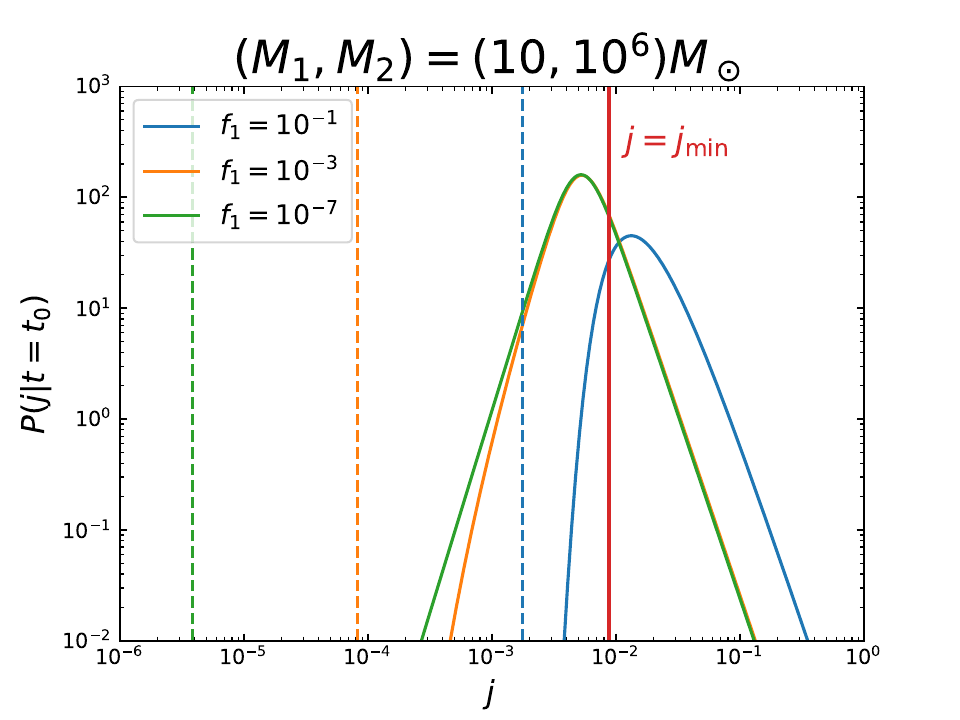}}
    { }
{\includegraphics[width=8cm]{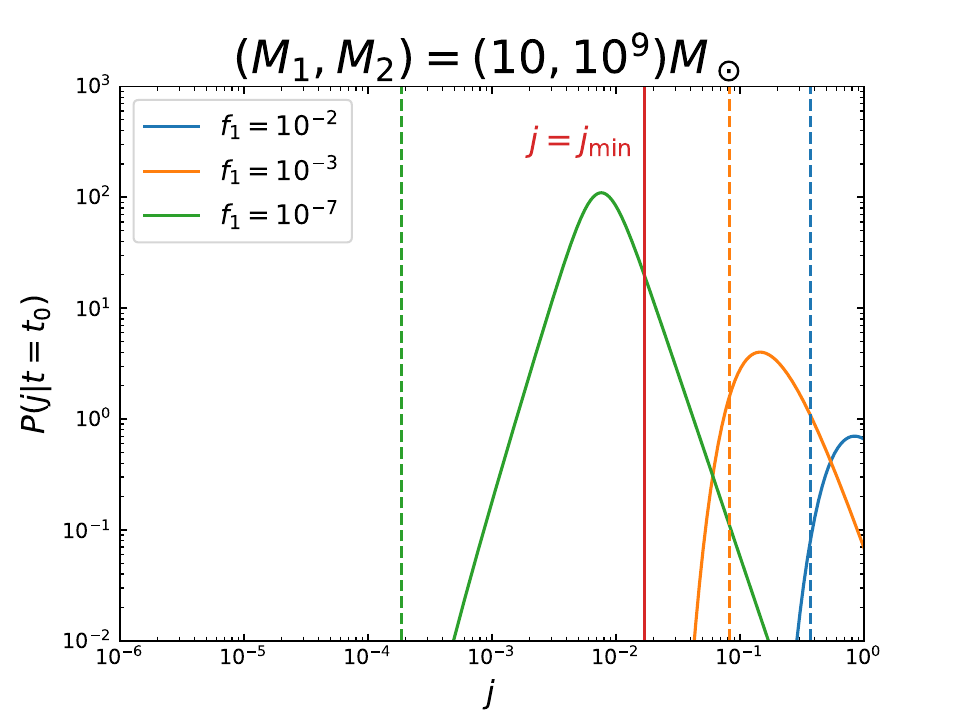}}
\caption{Distribution of initial $j$ for primordial EMRIs merging
today. We change the value of $f_1$ with fixed
$f_\text{PBH}=10^{-3}$. The threshold $j_c$ for which $R_p=R_s$ in
each case is indicated by the corresponding vertical line. }
\label{fig:rprs}
\end{figure}

In \autoref{fig:rprs}, we depict the distribution of $j$ for two
cases: $(M_1,M_2)=(10,10^6)M_\odot$ and $(10,10^{9})M_\odot$, both
having the same total energy density $f_\text{PBH}=10^{-3}$. It is
observed that both the probability distribution $P(j|t)$ and the
critical $j_c$ essentially remain unchanged for $f_1<10^{-5}$,
and we can safely disregard the effect of \textsf{direct plunge}
since the probability of $j>j_c$ for $j>j_{\text{min}}$ approaches
unity.


\section{Astrophysical EMRI models} \label{sec:appendixB}

\begin{table*}
\centering
\begin{tabular}{ccccccc|c}
\hline
Model & Mass function & MBH spin & Cusp erosion & $M$--$\sigma$ relation & $N_\mathrm{p}$ & CO mass [$M_\odot$] & EMRI rate [$\mathrm{yr}^{-1}$]   \\
\hline
M1 & Barausse12 & a98   & yes & Gultekin09    & 10  & 10 & 1600  \\ 
M2 & Barausse12 & a98   & yes & KormendyHo13  & 10  & 10 & 1400  \\ 
M3 & Barausse12 & a98   & yes & GrahamScott13 & 10  & 10 & 2770  \\ 
M4 & Barausse12 & a98   & yes & Gultekin09    & 10  & 30 &  520  \\ 
M5 & Gair10     & a98   & no  & Gultekin09    & 10  & 10 &  140   \\ 
M6 & Barausse12 & a98   & no  & Gultekin09    & 10  & 10 & 2080   \\ 
M7 & Barausse12 & a98   & yes & Gultekin09    & 0   & 10 & 15800   \\ 
M8 & Barausse12 & a98   & yes & Gultekin09    & 100 & 10 &  180   \\ 
M9 & Barausse12 & aflat & yes & Gultekin09    & 10  & 10 & 1530   \\ 
M10 & Barausse12 & a0    & yes & Gultekin09    & 10  & 10 & 1520  \\ 
M11 & Gair10     & a0    & no  & Gultekin09    & 100 & 10 &   13 \\ 
M12 & Barausse12 & a98   & no  & Gultekin09    & 0   & 10 & 20000 \\ 
\hline
\end{tabular}
\caption{List of astrophysical EMRI models considered in
Ref.~\cite{Babak:2017tow}. Column 1 defines the label of each
model. In other columns the following quantities are specified:
the MBH mass function (column 2), the MBH spin model (column 3),
whether or not the effect of cusp erosion is included (column 4),
the $M$--$\sigma$ relation (column 5), the ratio of plunges to
EMRIs (column 6), the mass of the compact objects (column 7), the
total intrinsic EMRI rates (yr$^{-1}$) up to $z=4.5$
(column 8).} \label{tab:tab1}
\end{table*}

\bibliography{refs}

\end{document}